\documentstyle[preprint,aps]{revtex}
\begin{document}
\tighten
\draft
\preprint{THU-95/15; To appear in Phys.~Rev.~B}
\title{Vortex reflection at boundaries of Josephson-junction
arrays.}
\author{T. J. Hagenaars$^{1,*}$, J. E. van Himbergen$^1$,
Jorge V. Jos\'{e}$^{2}$, and P. H. E. Tiesinga$^1$ \\
{\it $^1$Instituut voor Theoretische Fysica,\\ Princetonplein 5,
Postbus 80006, 3508 TA Utrecht, The Netherlands\\
$^2$Department of Physics, Northeastern University,\\ Boston
Massachusetts 02115, USA}}
\date{July 6, 1995}
\maketitle
\begin{abstract}
We study the propagation properties of a single vortex in
square Josephson-junction arrays (JJA) with free
boundaries and subjected to an applied dc current.
We model the dynamics of the JJA  by the resistively and capacitively
shunted junction (RCSJ) equations.
For zero Stewart-McCumber parameter $\beta_c$ we find
that the vortex always escapes from the array when it gets to the boundary.
For $\beta_c\geq 2.5$ and for low currents we find that the vortex escapes,
while
for larger currents the vortex is reflected
as an antivortex at one edge and the antivortex as a vortex
at the other,
leading to a stationary vortex oscillatory state and
to a non-zero time-averaged voltage.
The escape and the reflection of a vortex at the array edges is
qualitatively explained in terms of a coarse-grained model of a
vortex interacting logarithmically with its image.
For $\beta_c\geq 50$ we find  that the reflection regime is split up in
two disconnected regimes separated by a second vortex escape regime.
When considering an explicit
vortex-antivortex pair in an array with periodic
boundaries,
we find
a soliton-like
non-destructive collision
in virtually the same current regimes
as where we find reflection of a single vortex
at a free boundary;
outside these current regimes the pair annihilates.
We also discuss the case when the
free boundaries are at $45$ degrees with respect to
the current direction, and thus the angle of incidence of the
vortex to the boundaries is $45$ degrees.
Finally, we study the effect of self-induced magnetic fields (for
penetration depths ranging from 10 to 0.3 times the lattice
spacing)
by taking into account the full-range inductance
matrix of the
array and find qualitatively equivalent results.  We also discuss possible
consequences of these results to experimental systems.

\end{abstract}


\section{Introduction}
In recent years the viscous motion of a single vortex
in two-dimensional Josephson-junction arrays (JJA)
has been studied numerically by several
authors \cite{Bobb,UliG,YuS,YuS2,Hage,Hage2,UliG2}.
In these studies the I-V characteristics were calculated for JJA
with periodic boundary conditions perpendicular to the
direction of the applied dc current.
Apart from the vortex viscosity, the nature of its predicted mass
\cite{Simanek,Los,EckSch,Eck,Rzch,OrlM,EckSon} has attracted significant
interest,
also experimentally \cite{HHerre,Herre1}.
Van der Zant et al. reported experimental evidence
for ballistic motion of vortices in a current-free region in  a
highly underdamped Josephson-junction array \cite{HHerre}.

An interesting aspect of this problem, which is relevant to experiments
such as those in \cite{HHerre}, but that has not yet been studied
systematically, is the influence of boundaries on the
vortex motion.
Experimentally one may distinguish two types of boundaries (see e.g. Ref.
\cite{HHerre}):
either the junctions at the edge are connected to a
superconducting busbar or they are not (free boundary).
The influence of the boundary on a vortex due to the
busbar geometry can be described by an image vortex of the same sign,
equidistant on the other side of the boundary; in the free boundary case
the
image of the vortex is of opposite sign.
The latter situation is to some extent analogous to that of a soliton
in a continuous long Josephson junction (LJJ) \cite{FulDy,Naka,LaSco}.

The goal of this paper is to study the reflection, transmission and
annihilation properties of vortices in JJA with free boundaries.
Here, we report simulations of dc biased arrays with free
boundaries, in zero applied magnetic field.
For every bias current considered we use the same
initial phase configuration with one vortex in the middle
of the array.  We study the dynamical interaction of the vortex
with the free boundary as it moves towards it  for bias
currents above the depinning threshold.
The vortex-boundary interaction is equivalent to  the effect of
an image antivortex
at equal distance on the other side of the boundary.
A vortex and an antivortex interact logarithmically
(at sufficiently large distances), so the same holds for
a vortex and a free boundary.
The total effective vortex potential
is the sum of the interaction
potential and the periodic lattice pinning potential.
The Lorentz force on the vortex due to the applied bias
current corresponds to a tilt of this effective potential.
Below we will see this more explicitly when we
discuss the coarse-grained model equations for the vortex dynamics.

The outline of this paper is the following.  In section II we define the
model equations studied in this paper.  There we consider the extreme
type-II limit (infinite penetration depth)
as well as the case of finite penetration depth.  We also
discuss the coarse-grained vortex equations which are used to analyze the
results later in the paper.  In section III (a) we present results in the
extreme type-II regime and in III (b) the results for finite
magnetic penetration depth.
In
section IV we analyze the results from III (a) in terms of the
coarse-grained, phenomenological vortex equations including the interaction of
the vortex with the free boundaries.
 In section V we present results for a vortex moving in diagonal JJA.
Section VI contains a summary of our results and some conclusions.

\section{Model Equations}

In JJA vortices are represented by patterns of eddy current
around a plaquette.
We consider dc-biased JJA in
the classical regime defined by $E_{J}\gg E_{c}=e^2/2C$,
where $E_J$ is the Josephson coupling energy and $E_c$ the
characteristic charging energy of a junction, $e$ the electron charge,
and $C$ the mutual capacitance of the junctions.
In this case, and in zero applied magnetic field, the individual
junctions in the JJA can be
modeled by the resistively and
capacitively shunted junction (RCSJ) model \cite{Likh}, defined by the total
bond
current $i(\bbox{r},\bbox{r'})$ between nearest neighbor
sites $\bbox{r}$ and $\bbox{r'}$ as
\begin{equation}
i(\bbox{r},\bbox{r'})=
\beta_{c}\ddot{\theta}(\bbox{r},\bbox{r'})
+ \dot{\theta}(\bbox{r},\bbox{r'})
+\sin[\theta(\bbox{r},\bbox{r'})].\label{RSJ}
\end{equation}
Here the dots represent time derivatives. The three contributions
to  $i(\bbox{r}, \bbox{r'})$ are the  displacement, the dissipative and
the superconducting currents, respectively.
The phase difference across a junction is
$\theta(\bbox{r},\bbox{r'})\equiv
\theta(\bbox{r})-\theta(\bbox{r'})$.
The currents are expressed in
units of the junction critical current $I_{c}$;
time is measured in units of the characteristic time
$1/\omega_{c}=\hbar/(2eR_{n}I_{c})$,
$\beta_{c}=(\omega_{c}/\omega_{p})^2$ is the Stewart-McCumber
damping
parameter \cite{Likh}, with the plasma frequency $\omega_{p}$ defined as
$\omega_{p}^2=2eI_{c}/\hbar C$,  and $R_n$
is the junction's normal-state resistance.
The RCSJ model has a limitation. It does not take into account
that the resistance  of a real Josephson junction with high $\beta_c$
is voltage-dependent:
for voltages below the gap voltage the effective resistance
is determined by the quasi-particle resistance, which can be
orders of magnitude larger than the normal-state resistance \cite{HHerre}.

In Fig. \ref{fig1} we show the geometry of the square
array to which most
of the results discussed in this paper apply.
Each superconducting island is connected to four neighbor
islands
via identical Josephson junctions.
The bias current is fed in at the lower boundary and taken out
at the upper boundary.

When we neglect the self-induced magnetic field produced by the
current flowing in the array (infinite magnetic penetration
depth $\lambda$), the array dynamics is described
by Eq. (\ref{RSJ}), together
with  Kirchhoff's current conservation condition
\begin{equation}
\sum_{\bf{a}}
i({\bf r},{\bf r}+{\bf a})=i_{\mbox{ext}}({\bf r}),\label{curco}
\end{equation}
imposed at every island ${\bf r}$ (the summation is over all
nearest neighbor islands $\bf{r}+{\bf a}$).
We integrate the coupled equations (\ref{RSJ}) and
(\ref{curco}) using
a fast algorithm discussed in \cite{DickFFT}.

When we include the self-induced fields,
thus including screening effects,
we redefine
$\theta({\bf r},{\bf r'})$ in (\ref{RSJ}) to be
the gauge invariant phase difference across a junction:
$\theta({\bf r},{\bf r'})\equiv
\theta({\bf r})-\theta({\bf r'})-2\pi A(\bf{r},\bf{r'})$.
The bond frustration variable
$A(\bf{r},\bf{r'})$ is defined by the line integral of the vector
potential $\bf{A}$:
\begin{equation}
A({\bf r},{\bf r'})=\frac{1}{\Phi_{0}}
\int_{\bf{r}}^{\bf{r'}}\bf{A}\cdot d\bf{l},
\end{equation}
where $\Phi_0=h/2e$ is the elementary flux quantum.
The vector potential is time-dependent, as is the
total magnetic flux $\Phi({\bf R},t)$ at plaquette $\bf{R}$:
\begin{equation}
\frac{\Phi({\bf R},t)}{\Phi_0}=\sum_{{\cal P}(\bf{R})} A({\bf r},{\bf
r'},t).
\end{equation}
Here ${\cal P}(\bf{R})$ denotes an anti-clockwise sum around
the plaquette $\bf{R}$.
The time-dependent magnetic flux is
induced by all the currents flowing in the array via the
Faraday-Ampere laws.  In zero external magnetic field, we may write
\begin{equation}
\Phi({\bf R},t)=\sum_{{\bf r},{\bf a}}\Gamma
({\bf R},{\bf r},{\bf a})i({\bf r},{\bf r+a},t).\label{sepa}
\end{equation}
$\Gamma$ is a matrix that explicitly depends on the
geometry of the array and the junctions.
The standard inductance matrix is then given by the discrete curl of
$\Gamma$.
As discussed in more detail in Ref. \cite{DanJor},
in the linearized approximation the strength of the self-induced fields
is governed by the perpendicular magnetic penetration depth
\begin{equation}
\lambda=\frac{\Phi_0}{2\pi I_c\mu_0},
\end{equation}
or equivalently by the $\kappa$-parameter
\begin{equation}
\kappa = \frac{\lambda}{a}=\frac{\omega_\Phi}{\omega_c},
\end{equation}
where $\omega_\Phi =R_n/(\mu_0 a)$,  and $\mu_0$ is the vacuum magnetic
permeability.
We set the lattice constant $a$ equal to unity and use the
parameter $\lambda$ to indicate the degree of screening.

In our calculations we use the full-range inductance matrix
(``Model C" in Ref. \cite{DanJor}).
The inductance matrix elements are computed using the model
approximations discussed in Ref. \cite{DanJor}.
After choosing the temporal gauge, equations (\ref{RSJ}) to
(\ref{sepa}) can be combined into one (vector) equation that governs the
complete dynamics (see Ref. \cite{DanJor}).

The vorticity $n(\bf{R})$ of a plaquette $\bf{R}$ is defined as
\begin{equation}
2\pi n({\bf R})=2\pi \Phi({\bf R}) +\sum_{{\cal P}({\bf R})}
\big(\theta({\bf r})-\theta({\bf r'})-2\pi A(\bf{r,r'})\big).
\label{vortdef}
\end{equation}
Here the gauge invariant phase difference
$\theta({\bf r})-\theta({\bf r'})-2\pi A({\bf r},{\bf r'})$ is
taken between $-\pi$ and $+\pi$.

We would like to model
the vortex motion in the array by a coarse-grained
equation of motion as in
the Bardeen-Stephen model for flux flow in
continuous superconductors.
The dissipative currents in the junctions, modeled in
Eq. (\ref{RSJ}) by currents through ohmic shunt resistors,
give rise to a viscous
force on a moving vortex.
Furthermore, in capacitive arrays ($\beta_c >0$), the electromagnetic
energy
stored in  the  junction
capacitors due to the vortex motion, can be interpreted
as the kinetic energy of a massive vortex
\cite{Los,EckSch,Eck,Rzch,OrlM,EckSon,Dick}.
For an infinite array, and for infinite magnetic penetration
depth,
a  coarse-grained model equation
in terms of a single
continuous vortex
coordinate $x$ reads:
\begin{equation}
M\ddot{x}+\eta \dot{x}=i_{b}+i_{d}\sin(2\pi x),\label{eom1}
\end{equation}
where $M=\pi\beta_{c}$ and $\eta=\pi$ for
a square array
\cite{Los,EckSch,Eck,Rzch,OrlM,EckSon,Dick}.
This equation describes the vortex as a point
particle with mass $M$ that, driven by a (Lorentz) force proportional
to $i_{b}$,
moves through a sinusoidal
pinning potential and experiences a viscous damping force with constant
viscosity coefficient $\eta$.
An estimate for $i_d$ in
a square lattice gives $i_d\approx 0.10$\cite{LAT}.
In a previous paper \cite{Hage}, we have deduced an alternative
phenomenological vortex equation of motion
with a velocity-dependent vortex  viscosity.
In contrast to
the model with  constant viscosity given in Eq. (\ref{eom1}),
the nonlinear model gives a qualitatively correct account of the
numerical results
for the current-voltage characteristics calculated
using the full set of equations (\ref{RSJ}).  The equation reads

\begin{equation}
M(\beta_c)\ddot{x}+\frac {A(\beta_c)}{1+B(\beta_c)\dot{x}}\, \, \,
{\dot{x}}
=i_{b}+i_{d}\sin(2\pi x).\label{nonl}
\end{equation}
Here the phenomenological
constants $A$, $B$ and $M$ are found to be weakly dependent
on $\beta_c$.

In the model equations (\ref{eom1}) or (\ref{nonl}) the vortex dynamics
is described in terms of a single particle with coordinate $x$.  In
writing these equations we assume that, as the vortex moves, it does
not couple to other excitations in the array.
As a result, it is assumed that the electromagnetic energy associated with
the
moving vortex, and interpreted as its kinetic energy,
can only be absorbed by the viscous medium and not transmitted
to other dissipative modes in the array.
The limited validity of this assumption is
apparent for example from
numerical simulations \cite{Bobb,UliG,YuS2},
where no ballistic vortex motion was found when switching off the bias
current.  Furthermore,  it was found in experiments \cite{Herre1}  and
simulations
that the vortex viscosity does not decrease as $1/R_n$ when increasing
$R_n$.
The non-zero vortex viscosity in the underdamped limit
was understood as due to the coupling
of the moving vortex to charge oscillations (`spin waves')
\cite{UliG,UliG2,EckSon,Herre1}.

For a finite array, an extra position-dependent
force due to the vortex-boundary interaction should enter at the
 right-hand side of these equations:
\begin{equation}
M\ddot{x}+\eta \dot{x}=i_{b}+i_{d}\sin(2\pi x)-F(x),\label{eq:modelF}
\end{equation}
and
\begin{equation}
M(\beta_c)\ddot{x}+\frac {A(\beta_c)}{1+B(\beta_c)\dot{x}}\, \, \,
{\dot{x}}
=i_{b}+i_{d}\sin(2\pi x)-F(x).\label{eq:modelF2}
\end{equation}
To compute this force, one has to take into
account all the infinitely many image vortices that are produced by
the array boundaries. When we take the origin at the left-hand
side boundary of the array, this leads to the following form of the
force:
\begin{equation}
F(x)= -\frac{d}{dx}\mbox{ln}(\frac{2L}{\pi}\sin(\frac{\pi
x}{L}))\label{Force}.
\end{equation}
In order to avoid the singularities at the boundaries in (\ref{Force}),
we use a cutoff: for distances to the boundaries
smaller than half a lattice constant, we set $F(x)=0$.
Note that this cutoff prescription fixes the maximum
attractive force due to the boundary on the vortex at
$F_{\mbox{max}}=F(\mbox{$x=\frac{1}{2}$})$.
In Ref. \cite{OrlM} it is argued that the vortex mass
increases close to a free boundary.
It was found that this increase is only noticeable within
one lattice constant from the boundary.
We do not include this quantitative correction in the models
(\ref{eq:modelF}) and (\ref{eq:modelF2}),
as the arbitrariness of the cutoff prescription already
makes
that the results are only qualitative in nature.

\section{Results }
\subsection{$\lambda=\infty$ }

In Fig. \ref{fig2}  we present a summary of the results for a
$16\times 16$ square array with
$\lambda=\infty$ for different values of $\beta_c$.
At the bottom we show the results of the simulations
for $\beta_c=0$,
in which the vortex motion is purely
viscous,
and the vortex mass is zero.
We have considered all current values between $i_b=0.0$ and
$i_b=1.0$, with a grid of $\Delta i=0.01$.
For this system size the vortex depinning current is between
$0.11$ and $0.12$, so for
currents  $i_b\leq 0.11$ the vortex is pinned to
the middle plaquette of the array, and for $i_b\geq 0.12$ it is depinned
(If one increases the system size the depinning current will
reach the value $i_d=0.10$ estimated in Ref. \cite{LAT}
for an infinite array).
When the vortex is depinned and moves towards one of the
free boundaries,  the approach to its image on the
other side of the boundary causes it
to accelerate.
When the vortex reaches the boundary, it
leaves the array, or, equivalently, it is annihilated by its
image antivortex. We denote this behavior as type {\bf A}.

Above $i_b=0.96$ the $\beta_c=0$ dynamics looses its
single-vortex character after a short transient time,
when additional vortices are generated in other rows.
In fact the measured voltage
is due to the motion of vortices in all rows.
As we want to focus on the single-vortex dynamics in this
paper, we will not go further into this type of motion here.

For nonzero $\beta_c$,
we started the simulations at $t=0$
increasing the current linearly
from zero  to
$i_b$ in one $RC$ time of the junctions.
The relaxational oscillations due to this increase in the current
decay while the vortex is moving towards the boundary.
We checked that the behavior at the boundary is
not seriously affected by these relaxational
oscillations
by comparing the results to those in a $32\times 16$ array.

For $\beta_c=2$, the results for the vortex propagation
 are similar to the $\beta_c=0$
results. The only difference is
that the regime with type {\bf A} behavior now ends at $i=0.86$
due to the onset of row switching, which means that the
vortex switches one or more rows of longitudinal junctions
(i.e. junctions in the current direction)
into the resistive state \cite{Bobb,UliG,YuS,HerreR}.

For $\beta_c\geq 2.5$ a current range opens up in which
the vortex is reflected as an antivortex at the
boundary in a way a soliton reflects in a long Josephson-junction (LJJ)
\cite{FulDy,Naka}.
Alternatively, one may say that the vortex and its antivortex image
pass through each other, analogously to
the non-destructive soliton-antisoliton collisions in
a LJJ \cite{FulDy,LaSco}.
The antivortex in turn is reflected at the opposite
free boundary as a vortex. This sequence repeats so that
the vortex/antivortex
never escapes from the array, thus producing
a non-zero time-averaged voltage
perpendicular to the vortex motion.
We denote this as type {\bf B} behavior.
In a LJJ a similar type
of soliton motion
gives rise to the first zero-field step \cite{Likh}.
Although there are important differences between
the properties of solitons in a continuous junction and
the properties of vortices in the discrete array,
it appears, in the regime considered here, that their reflection properties at
a
boundary as well as their collision properties are
similar in many respects.
For a two-dimensional array of Josephson junctions the  non-destructive
collision of a vortex-antivortex pair
has  been alluded to by Nakajima and
Sawada, in the context of a model including the self-inductances
of the plaquettes \cite{NaSa}.

In Fig. \ref{fig3} we show the voltage versus current characteristic for
$\beta_c=10$.  We note that for the calculation of this
current-voltage characteristic we use the
same initial phase configuration for all the bias currents considered.
Above the vortex depinning
current, we start the calculation of the time-averaged voltage
only after a  sufficiently long time interval for the vortex
to reach the boundary. As a result,
 in the case of type {\bf A} behavior
we measure no voltage, as there is no dissipation
after the vortex escapes from the array.
In contrast, the type {\bf B} regime has a non-zero
voltage. The current-voltage characteristic has
a considerable slope in this regime, in contrast to
the first zero-field step in a LJJ, which is much flatter,
due to the fact that the voltage is limited by the
maximum soliton velocity.
For even higher currents, we enter the row-switching regime
with a large voltage increase.

We interpret the type {\bf B} behavior as being the result of
the inertia, or kinetic energy,  carried by the vortex.
The attractive interaction between the vortex and its image
provides a potential well from which the
reflected antivortex has to escape after the
reflection/collision, in order to travel towards the
opposite boundary.
The Lorentz force on the antivortex due to the applied bias
current is in itself not sufficient to pull it out of the well.
In addition, the vortex needs to have a minimum kinetic energy
in order to escape:
this will be the case if both the vortex
mass $M=\pi\beta_c$ as defined
in equation (\ref{eom1}) and the vortex velocity (monotonically
increasing with  $i_b$) are sufficiently large.
In Fig. \ref{fig4} we show snapshots of the vortex configurations
at different times for $\beta_c=10$ and $i_b=0.49$.
In Fig. \ref{fig5} we show the time-dependent voltage across the array
indicating the times at which the snapshots shown in Fig. \ref{fig4}
are taken.
The large voltage fluctuations for short times are
due to the additive contributions of the
relaxational oscillations of all the individual longitudinal
junctions in the array as the bias current is changed
from zero to $0.49$ in the
time interval $0<t<10$.
For $t>100$ these oscillations have relaxed sufficiently and
the vortex contribution to the voltage becomes dominant.
As the vortex approaches the boundary its velocity increases and so
does the voltage. At the reflection, the vortex velocity jumps  from a
maximal value (just before the reflection) to a minimal value (just
after the reflection).

To verify the interpretation of the vortex reflection as a
non-destructive collision with the image antivortex,
we have investigated the collision properties of a
vortex-antivortex pair explicitly.
This entails   simulations of
a $16\times 32$ square array with {\em periodic
boundary conditions} in the direction of the vortex motion,
with one vortex and one antivortex separated by 16 lattice constants
present in the initial phase
configuration.
This situation is the explicit realization of
the image-vortex system
corresponding to the $16\times 16$
array with free boundaries and a single vortex.
Below the
row-switching threshold, we find that indeed the corresponding
vortex dynamics, i.e. destructive collision
{\bf A'} and  non-destructive collision {\bf B'},
takes place. The transition between the
types {\bf A'} and {\bf B'} behavior  is at a current value close to the one
between types {\bf A} and {\bf B} in the finite array with one vortex,
thus providing a phenomenological explanation of the results.

For currents in the type {\bf A'} regime close to {\bf B'},
the destructive collisions
become a two-stage process: first the
vortices collide non-destructively, drift three lattice
constants apart and then
fall back onto each other and annihilate.
In the LJJ-soliton language this is called a decaying
breather mode.
In the finite array the corresponding process in the {\bf A} range
(near to the {\bf B} range)  consists of a reflected antivortex
that falls back over the edge.
We interpret this process as evidence for
a non-zero vortex mass:
the vortex moves up-hill (in the potential landscape)
until it reaches a turning point where the
kinetic energy is used up, and then falls back in the potential
well.

When  $\beta_c$ increases, the lowest current resulting in
type {\bf B} behavior decreases, as shown in Fig. \ref{fig2}.
The current for row-switching
decreases as well, similar to the
$\beta_c$ dependence of the row-switching threshold in
previous simulations with periodic boundary conditions
\cite{Bobb,UliG,YuS,Hage}.
In the next section we will interpret the type {\bf A} versus type
{\bf B} behavior in terms of the
coarse-grained model equations given in Eqs. (\ref{eq:modelF}) to (\ref{Force})
of a
vortex interacting logarithmically with its image(s).

For $\beta_c \geq 50$ we observe the sequence
{\bf  A} $\to$ {\bf B} $\to$ {\bf A} $\to$ {\bf B}
 as a function of bias current:
the type {\bf B} regime
splits up into two pieces separated by a second type
{\bf A} regime.
The second type {\bf A} regime is due to the presence of
charge oscillations on the shunt capacitors,
which become larger in amplitude for increasing $\beta_c$.
At the reflection, the vortex velocity becomes so large,
that additional dissipative modes in the array are excited
in the form of local charge oscillations on the shunt capacitors.
These oscillations interfere with the motion of
the antivortex, which as a result
is slowed down and cannot escape from the potential well.
For higher currents, there is a second {\bf B} region.
Here the antivortex has a higher velocity, and it is
able to survive the coupling to the
charge fluctuations.
The magnitude of the second {\bf A} regime
grows with $\beta_c$, and we have found that for
$\beta_c=500$ there are  no
{\bf B} regimes left.
We note that the break up of the type {\bf B} regime into
two pieces and the
disappearance of the
reflection for very low damping have no counterpart in
the context of a  soliton moving in a LJJ.

One might ask if the collision can still be
non-destructive if the vortices collide along a direction that
is not parallel to one of the coordinate axes.
We have simulated a finite $16\times 16$ square array with
$\beta_c=10$ containing
a vortex and an antivortex that are inserted not in the same
row but in adjacent rows.  The results show a reduced
but non-vanishing current regime
($0.63\leq i_b\leq 0.66$) for reflections
for both vortices at the
edges plus a `new' non-destructive collision in the middle of
the array, during which the vortices interchange rows,
as shown in Fig. \ref{fig6}.

 \subsection{Finite $\lambda$ regime}

Thus far we have studied the reflection properties of the
vortex in an array with zero self-induced magnetic
fields, or equivalently, with magnetic penetration depth $\lambda$
much larger than the array size.  From the
point of view of for example arrays made of niobium junctions,
in which the self-induced magnetic fields are
non-negligible \cite{Phillips}, it is important to see how these
fields influence the reflection properties of a vortex.
We have therefore studied the vortex propagation in
$16\times 16$ square arrays for a range of  $\lambda$
values at $\beta_c=10$ and $100$.
We find that the reflection still occurs for
finite $\lambda$.
The results are shown in Figs. \ref{fig7} and \ref{fig8}.
As discussed by Phillips et al. \cite{Phillips},
the vortex depinning current is enhanced by the self-induced
fields. For example, for our $16\times 16$ array with $\lambda =0.3$
and $\beta_c=100$ we find that the vortex depins for currents larger than
or equal to $i_b =0.38$.
 For $\beta_c=10$ and in the type-II regime
($\lambda$ larger than a few lattice spacings)
the width and the position
of the {\bf B} regime is somewhat insensitive to the $\lambda$ value
as one can see in Fig. \ref{fig7}.
As we see in Fig. \ref{fig8} for $\beta_c=100$, the
second {\bf A} regime shrinks with decreasing $\lambda$,
and disappears for $\lambda < 5$.
For $\lambda\leq 1$  (type-I)   we find that the
{\bf B} regime shrinks, both for $\beta_c=10$ and $\beta_c=100$.

\section{Comparison with the phenomenological model equations}

In this section we will make a qualitative comparison
of the $\lambda=\infty$ results presented in section III
with the results for the vortex motion in a finite array
based on Eqs. (\ref{eq:modelF}) and (\ref{eq:modelF2}).

Naively, in Eq. (\ref{Force}) the parameter range for $x$
is $0\leq x\leq L$. However, when the
vortex reflects at $x=L$, the positive image vortex of
the reflected antivortex has a coordinate $x >L$, and
we can interpret the oscillating state (type {\bf B} behavior)
as the motion of a positive vortex through the force field $F(x)$,
periodically extended to the complete real axis.
{\bf A} behavior then corresponds to trapping of the vortex
around $x=L$ (modulo $L$).
Starting from $x=8$ at each current value considered,
we calculated the current-voltage characteristic numerically
from
Eqs. (\ref{eq:modelF}) and (\ref{Force})
for $L=15$, using the cutoff prescription mentioned above.
The results for $\beta_c=10$  are
shown as the dashed line in Fig. \ref{fig3}.
The model
underestimates the threshold for type {\bf B} dynamics.
As  this threshold is very sensitive to the type of
cutoff used, we can not draw quantitative conclusions
about the vortex mass from this result.

We have determined the threshold currents
$i_{A\rightarrow B}$
as a function of $\beta_c$ from the model equation (\ref{eq:modelF}),
as well as from the same equation with the nonlinear friction
term from Eq. (\ref{eq:modelF2}).
We find that the dependence
of the $A\to B$ threshold current on $\beta_c$ is
qualitatively the same as found in the simulations.
In the models the decrease of the threshold value is due
to the increase in the mass parameter $M=2\pi\beta_c$, which
increases the vortex kinetic energy.
The larger this kinetic energy is, the easier it is for the vortex
to escape from the effective potential well.

We note that the way in which vortex inertia
manifests itself, and hence the (possible) attribution of
a mass,
depends on the dynamical situation considered.
In previous simulations
\cite{Bobb,UliG,YuS2},
in contrast to this work,
the vortex mass was probed by switching off the bias
current.
Changing the current gives rise to enhanced
local relaxational
oscillations in the
junction phases near to the vortex center, or
in other words, to a coupling to other dissipative
modes in the array.
Instead of allowing the vortex to continue its motion,
the electromagnetic energy stored in the capacitors then leads to
local oscillations. For large $\beta_c$ these force the vortex
center to oscillate back and
forth a couple of times between two adjacent plaquettes,
as shown for $\beta_c=2500$ in Fig. 10 of Ref. \cite{UliG}.
During the decay of these oscillations the electromagnetic energy is
dissipated.
The distance traveled by the vortex after the current has been
turned off, is zero. Therefore the mass attributed to the vortex in
this situation is zero or very small.
Similarly, probing the vortex mass by looking at the hysteresis in the
I-V characteristics \cite{YuS,Hage}, which also involves changes
in the current bias, leads to zero or very small
phenomenological mass $M(\beta_c)$ in (\ref{nonl})
for moderate values of $\beta_c$ \cite{Hage}.

On the basis of the model interpretation,
one would expect that, for some currents just below
the transition to type {\bf B} behavior,
the reflected
antivortex is not falling back over the boundary, but it is
re-trapped by the lattice.
This re-trapping  is then the effect of the combination of
the pinning modulation of the effective potential
and the energy loss due to  friction.
We have looked for  such a re-trapping process in the
simulations for $\beta_c=100$, and indeed found it for
$i_b=0.2462$ ($16\times 16$ array). The reflected antivortex is re-trapped
in the second plaquette from the boundary.

\section{Vortex reflection in diagonal arrays}

An important question is how robust the vortex reflection at a boundary
(or the non-destructive collision  of a vortex-antivortex pair)
is for other array geometries.
We have studied the case in which the current
is injected not along one of the coordinate axes, but along the $-x$ and
$y$ direction.
See Fig. \ref{fig9}, that shows a $15\times 15$  diagonal array,
in which the initial position of the vortex
is indicated by an open circle.
The vortex will move at
45 degrees to the boundary, due to the diagonal current bias,
which allows us to study the reflection behavior
for this angle of incidence.
We also find soliton-like reflection in this case, albeit the
width of the current regime is smaller, and vanishes again
already at a  $\beta_c$ value somewhere
between 50 and 100.
Although the vortex and its
image
collide at an angle of 90 degrees, the reflecting antivortex still
retraces the path of the incident vortex.
This is  basically because the
antivortex has to travel at right angles to the applied bias
current.
 An interesting question would then be what happens
for a   reflection at zero current bias.
To study this within the setting of our model simulation,
one would need a nonzero current first to depin the vortex
and then switch the current off just before the reflection.
However, it is known \cite{Bobb,UliG,YuS2}
that switching off the bias current
is not followed by a propagation of the vortex
for $\beta_c$ in the range in which we find vortex reflection.
Therefore, switching off the bias current just before the
reflection does not seem to be an option to
study reflection at zero bias current.

\section{Summary and discussion}

In conclusion, we have performed RCSJ-model simulations of
a finite Josephson-junction array containing a single vortex
in a dc current bias.
For moderate damping, we  find current regimes
where the vortex reflects at the boundary, moving back as an
antivortex towards the opposite boundary, where the
dual reflection process takes place, leading to a stationary
oscillatory state and a non-zero time-averaged voltage across
the array.
This oscillatory motion can be viewed as
a discrete analogue of the
bouncing of a soliton in an underdamped long Josephson
junction, which gives rise to the first zero-field step \cite{FulDy}.
The long Josephson junction or Josephson transmission
line can be used as a vortex flow
transistor or oscillator in many different devices \cite{Peder}.
Recently, Van der Zant and Orlando \cite{ZaOr}
explored the possibility of using a discrete one-dimensional parallel
array of underdamped junctions
as a vortex flow transistor.
Some time ago it has been shown numerically \cite{NaOn}
as well as experimentally \cite{Fuji,Naka2} that
vortices in a 1D parallel Josephson array have
soliton-like collision properties.
Our results show that the soliton-like collision and reflection also
occur in two-dimensional arrays,
and that the first zero-field step has a
discrete analogue in JJA,
which suggests a possible use of the two-dimensional array
as a vortex-flow device. An important difference found with respect to
the soliton motion in LJJ for very low damping, is the fact, that
in the array the reflection
is absent for values of $\beta_c$ of the order of 500 and larger \cite{LaSco}.

When considering a vortex-antivortex pair in an array with periodic
conditions,
the vortices collide non-destructively for appropriate bias
currents.
This is the behavior
equivalent to the reflection of one vortex at a free boundary.
Our numerical
results can be interpreted in terms of a macroscopic model
equation for a massive vortex.
We also studied
the vortex propagation in a diagonal array where the vortex
has an angle of incidence of  45  degrees to the boundary,
 and found that, for appropriate bias currents,
 reflection takes place  here as well.
Finally, we also studied the effect of self-induced magnetic fields,
using a model that  takes into account the
full inductance matrix of the array and found that the same type
of vortex reflection regimes as in the $\lambda=\infty$ case
are present for a range of values of $\lambda$ and $\beta_c$.

\section*{Acknowledgments}

We thank J.E. Mooij for suggesting the topic of this work.
We also thank him and W. Elion, P. Hadley, A. van Oudenaarden
and H. van der Zant for discussions.
This work was
supported in part by the Dutch organization for fundamental
research (FOM).
The work of JVJ has
been partially supported by NSF grant No. DMR-9521845.\\[1.0cm]
$^*$ Present address: Institut f\"ur Theoretische Physik,
Universit\"at W\"urzburg, Am Hubland, 97074 W\"urzburg, Germany.

\begin{figure}
\caption{Square array geometry used in the simulations, illustrated
with a $8\times 8$ array. Junctions are denoted as crossed bonds.
Free boundary conditions are imposed in both directions,
 while the current bias is applied  along the $y$-direction.}
\label{fig1}
\end{figure}

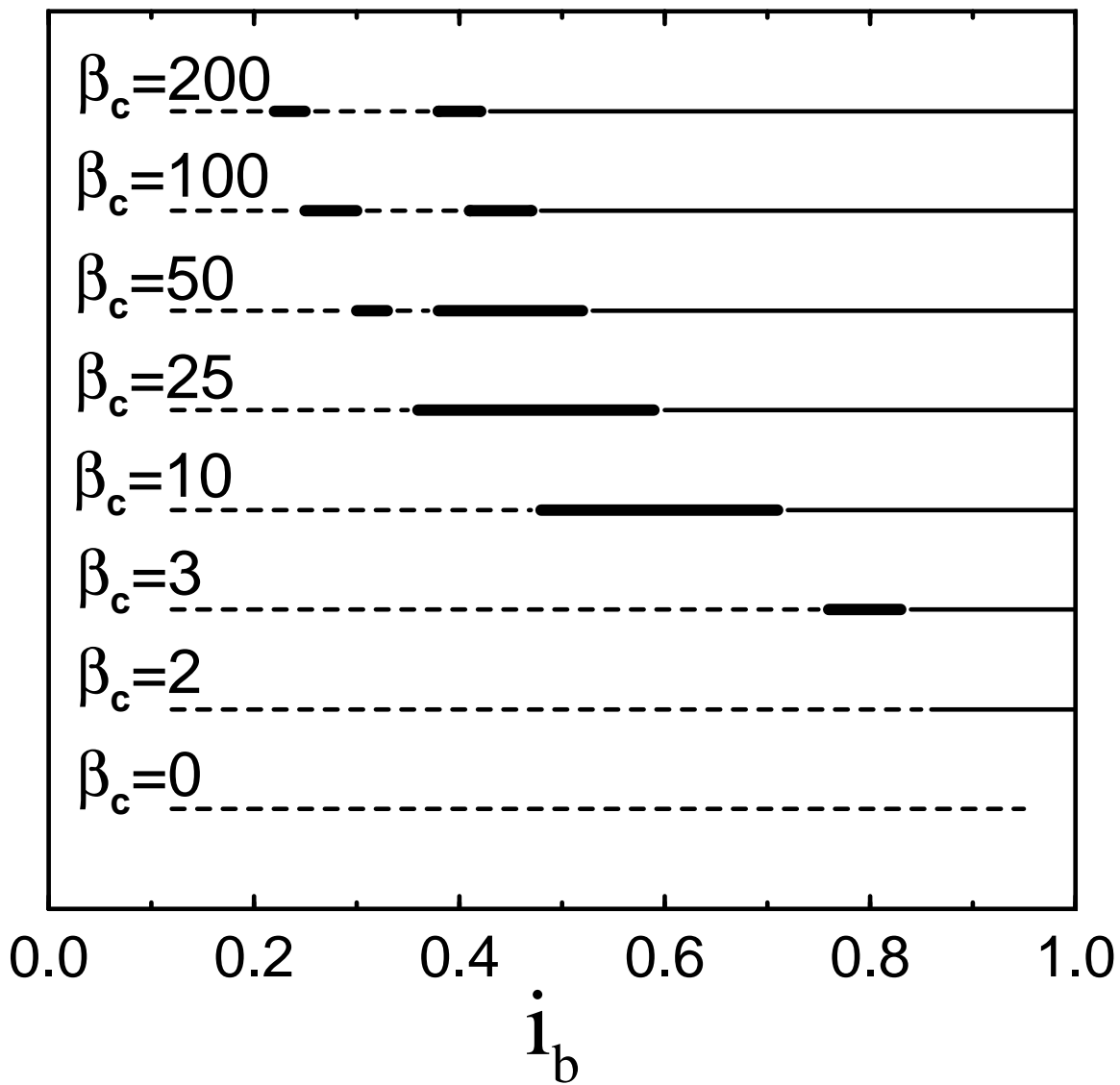
\begin{figure}
\caption{ Results for different values of $\beta_c$
for a 16$\times$16 square array with free boundaries.
The dashed lines represent the current ranges
for which type {\bf A} behavior (vortex escapes from array) was found.
The thick lines denote the type {\bf B} ranges (trapped vortex
oscillating in the array) and the thin full lines the row-switching
regimes.}
\label{fig2}
\end{figure}
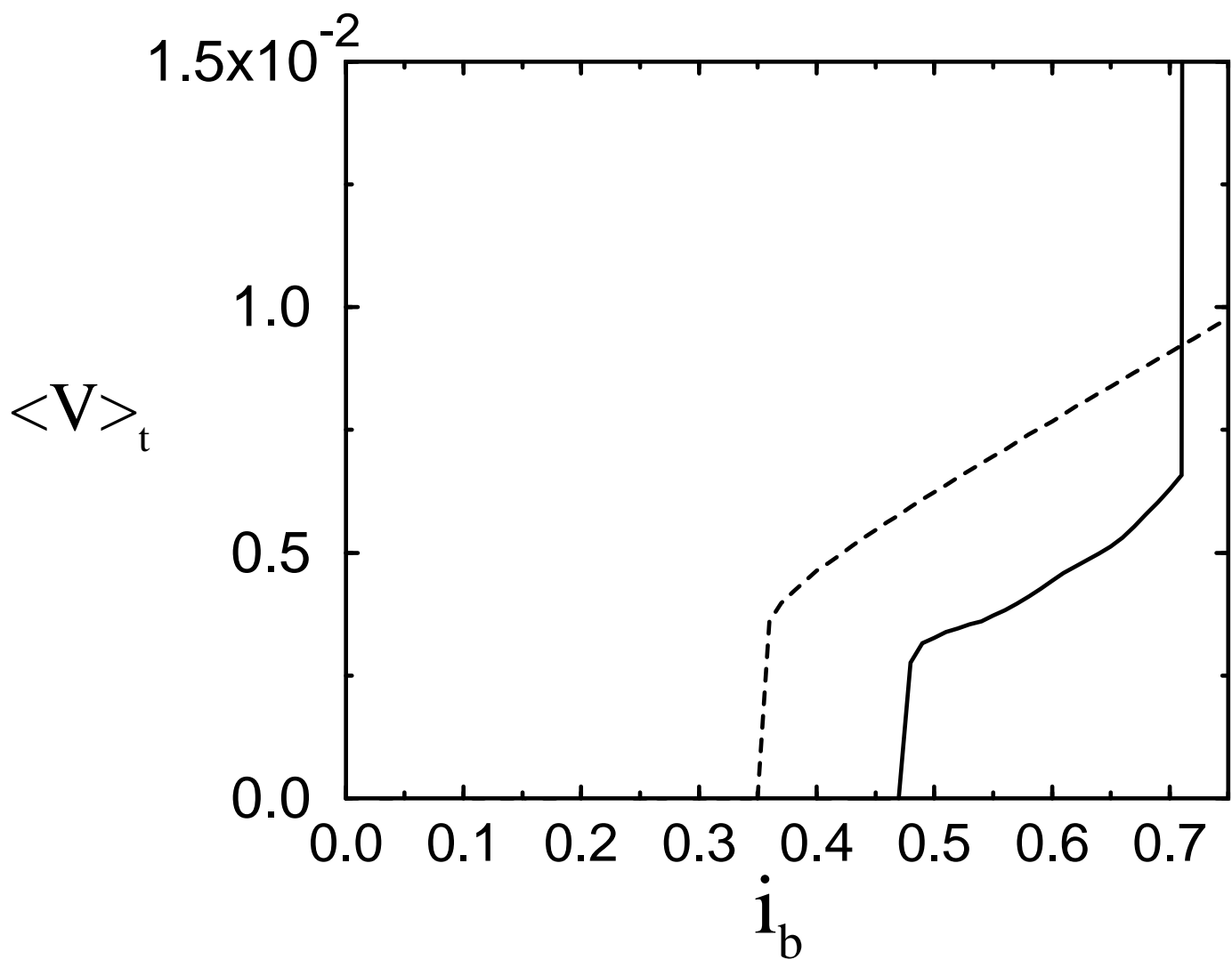
\begin{figure}
\caption{Current-voltage characteristics for $\beta_c=10$,
 from simulations of a $16\times 16$ square array with one
vortex (full line).
$\langle V\rangle_t$ is the time-averaged voltage across the array in the
current
direction, in units of $I_cR_n$ and normalized by the number
of longitudinal junctions. The dashed line is the I-V
characteristic computed from the linear viscosity model (11).}

\label{fig3}
\end{figure}
\begin{figure}
\caption{Snapshots of the vorticity distribution Eq.
(8) in a
$16\times 16$ square array with $\beta_c=10$ and $i_b=0.49$,
 showing type {\bf B} dynamics. In the frame labeled
with $0$ the vortex (black square $\equiv$ vortex center plaquette)
is moving towards the right free boundary.
After reflection it travels as an antivortex (white square)
towards the opposite boundary (frames 1 and 2), where it is
reflected again as a positive vortex (frame 3).
The snapshots correspond to the instants indicated in Fig. 5.}
\label{fig4}
\end{figure}
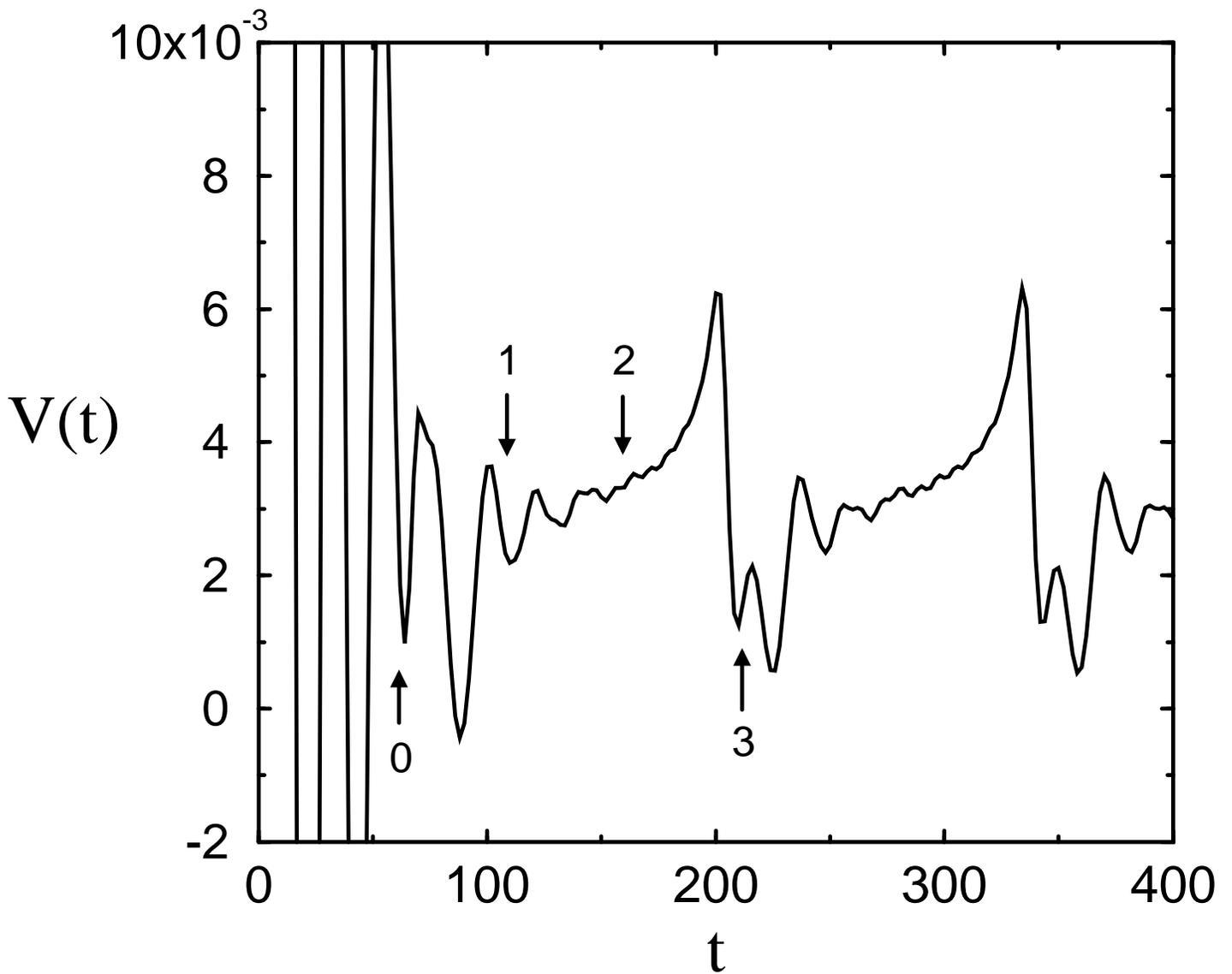
\begin{figure}
\caption{Voltage, normalized as in Fig. 3,
 versus time across a $16\times 16$ square array
with one vortex, for $\beta_c=10$
and $i=0.49$. The times labeled with 0 to 3 correspond
to the respective snapshots shown in Fig. 4.}
\label{fig5}
\end{figure}
\begin{figure}
\caption{Snapshots of the vorticity distribution
for a $16 \times 16$ array with free boundaries and
$i_b=0.65$, with one positive
vortex and one antivortex moving in adjacent rows.
The notation is as in Fig. 4.
Both vortices reflect at the edges (between frames 0 and 1),
and collide constructively in the middle of the array (between 2 and 3),
interchanging rows.}
\label{fig6}
\end{figure}

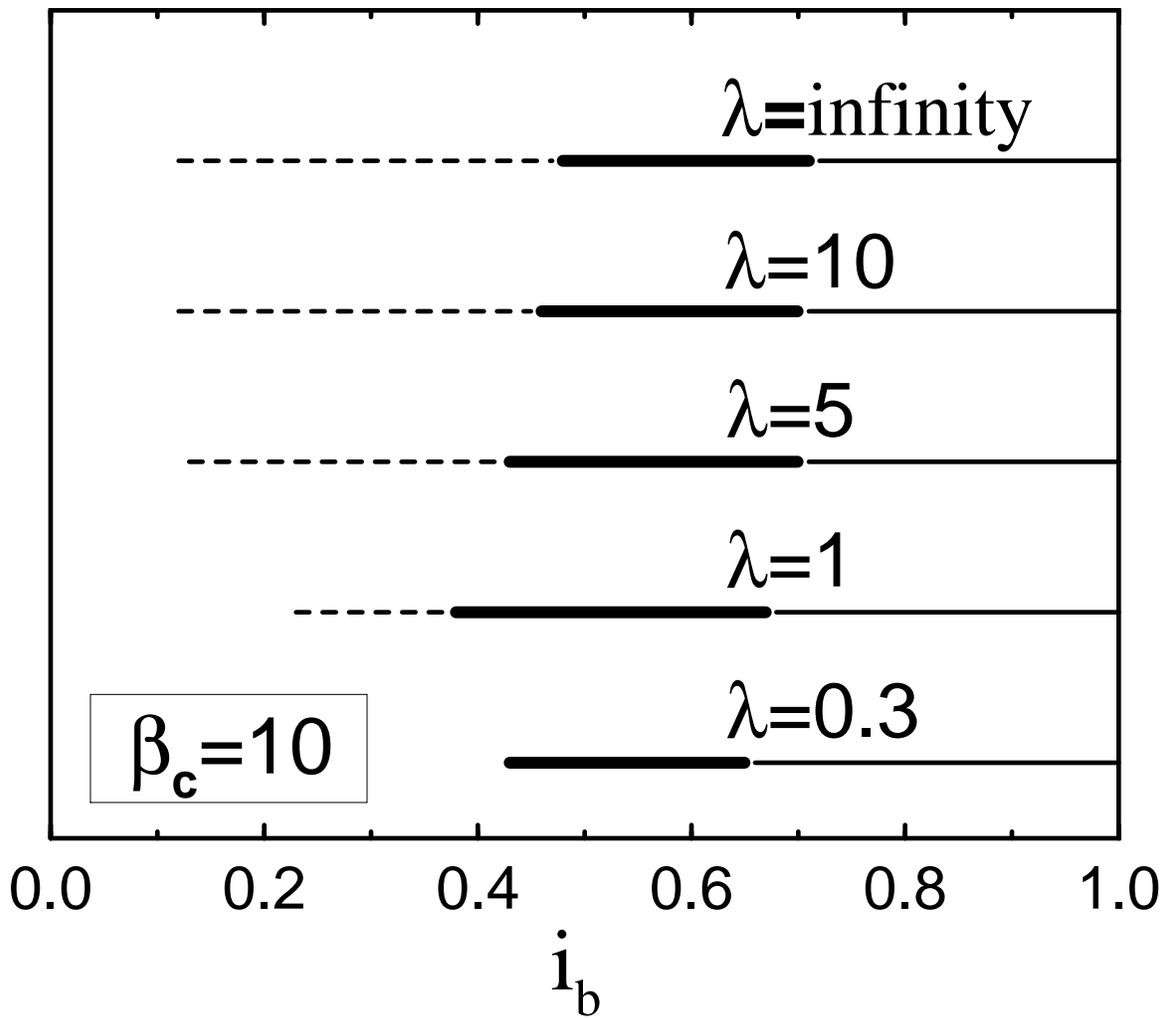
\begin{figure}
\caption{ Results for different values of $\lambda$
for a 16$\times$16 square array with free boundaries at
$\beta_c=10$.
The notation is the same as in Fig. 2. }
\label{fig7}
\end{figure}
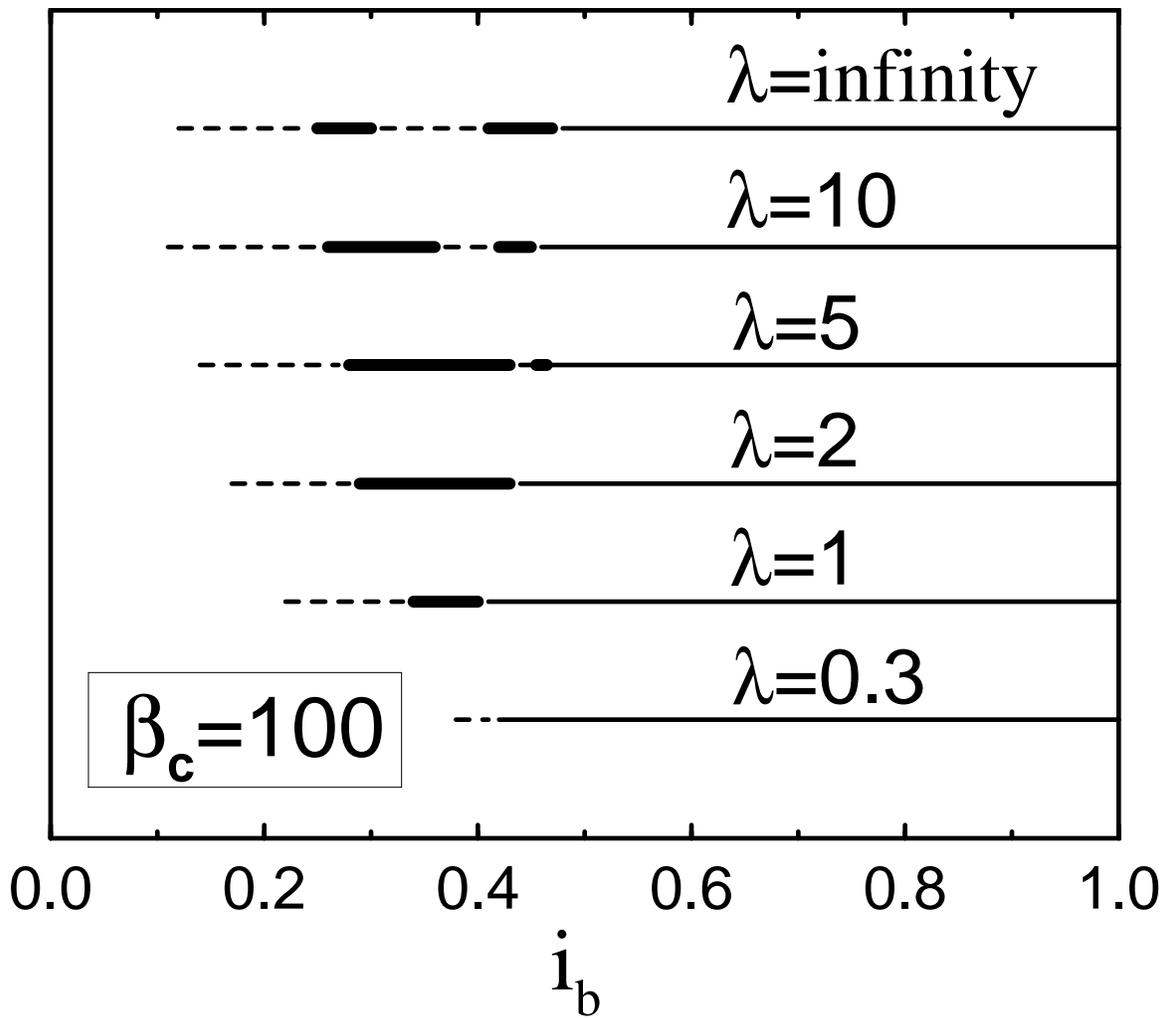
\begin{figure}
\caption { Results for different values of $\lambda$
for a 16$\times$16 square array with free boundaries at
$\beta_c=100$.
The notation is the same as in Fig. 2. }
\label{fig8}
\end{figure}

\begin{figure}
\caption{Diagonal array geometry used in the simulations, illustrated
with a $15\times 15$ array. For clarity we omitted the crosses
on the bonds.
In both directions free boundary conditions are
imposed, while the current bias is applied  along a diagonal
direction. The open circle denotes the initial position of
the vortex.}
\label{fig9}
\end{figure}
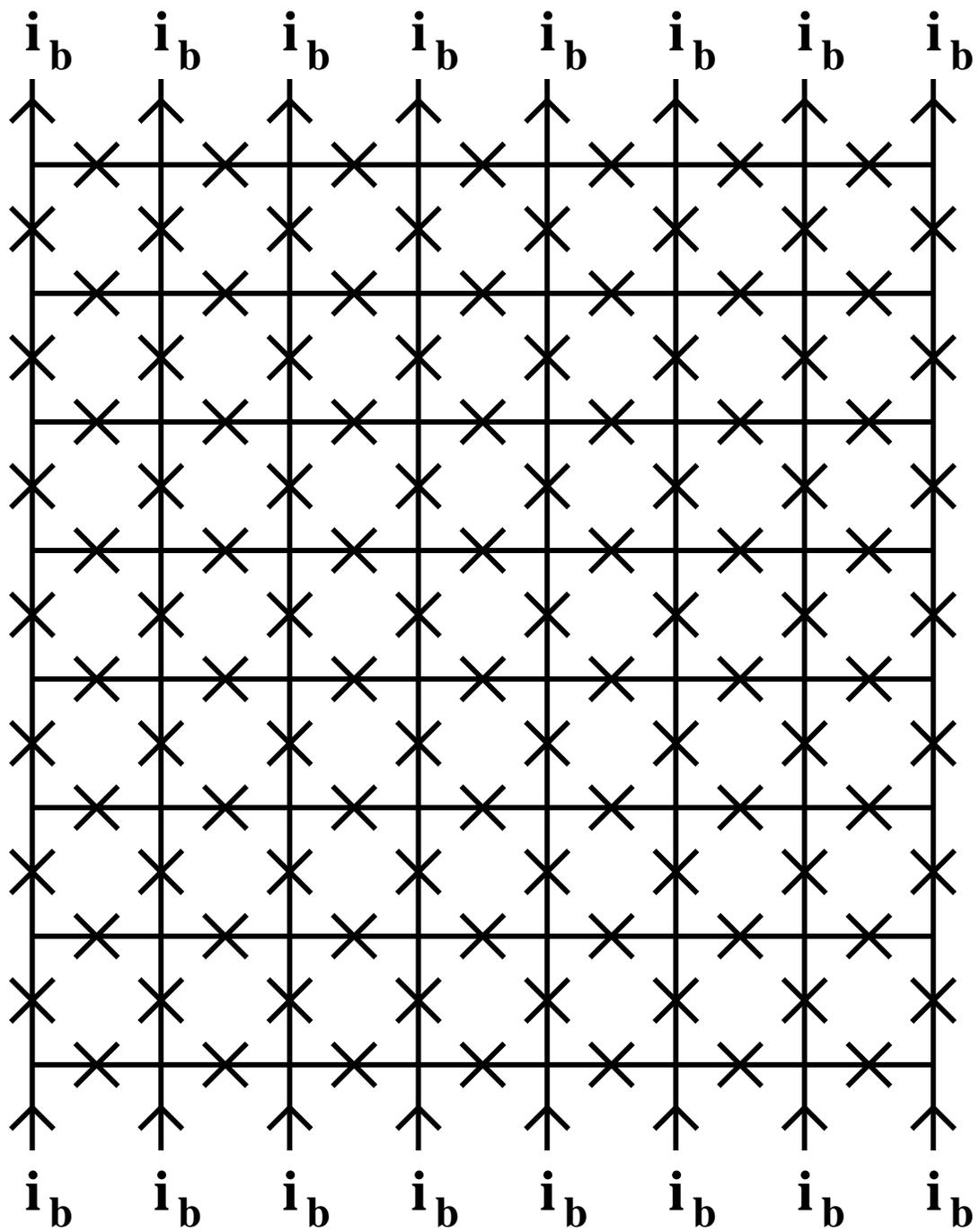
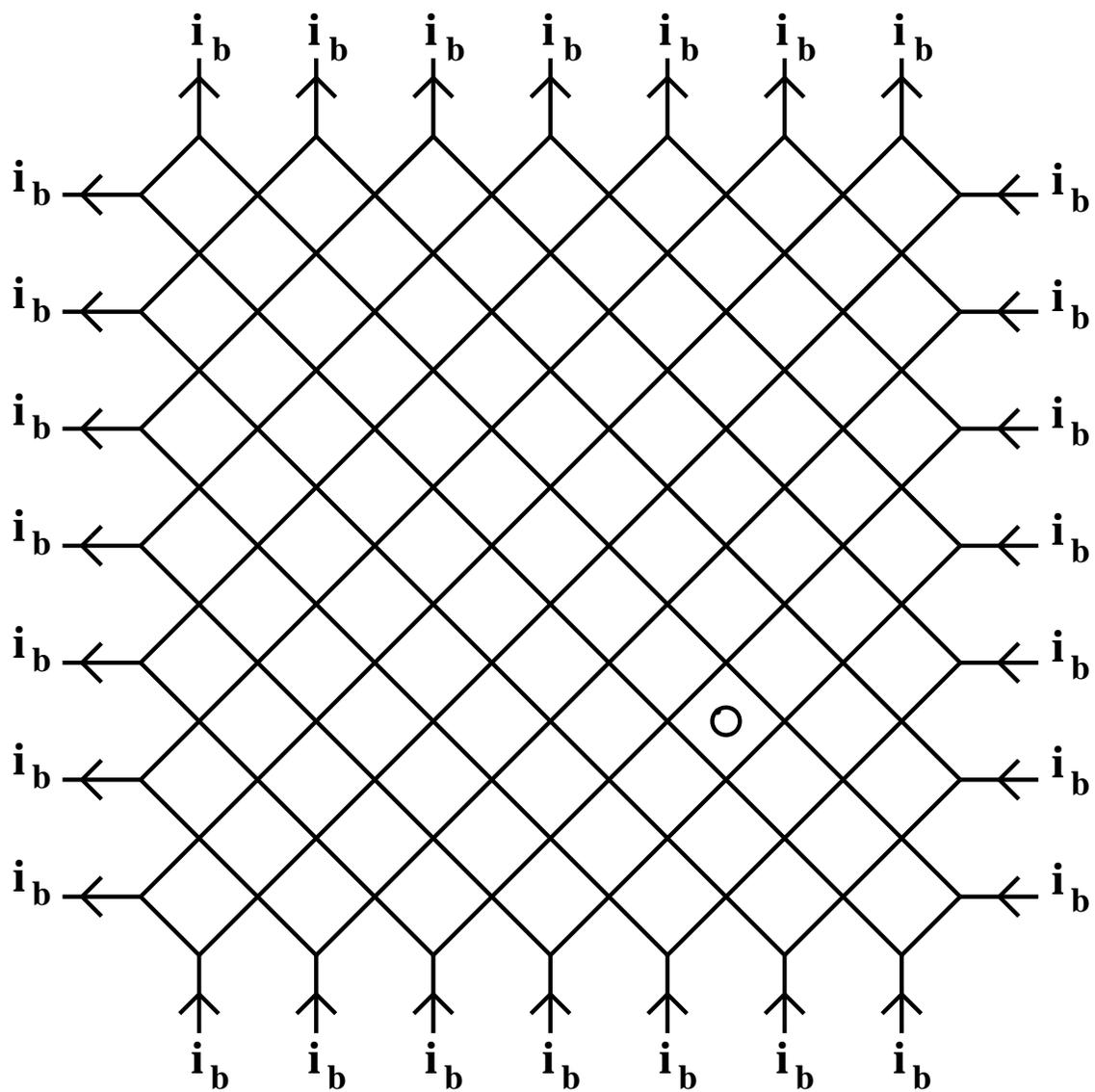
\end{document}